\documentclass{PoS}

\newcommand{\SU}{\mathrm{SU}}

\newcommand{\dd}{{\rm{d}}}
\newcommand{\Tr}{{\rm Tr\,}}

\newcommand{\SW}{S_{\mbox{\tiny{W}}}}

\newcommand{\THagedorn}{T_{\mbox{\tiny{H}}}}
\newcommand{\TNG}{T_{\mbox{\tiny{NG}}}}
\newcommand{\mthreshold}{m_{\mbox{\tiny{th}}}}

\newcommand{\Up}{U_{\mbox{\tiny{p}}}}

\newcommand{\npar}{n_{\mbox{\tiny{par}}}}

\newcommand{\eq}{\begin{equation}}
\newcommand{\en}{\end{equation}}
\newcommand{\eqar}{\begin{eqnarray}}
\newcommand{\enar}{\end{eqnarray}}

\title{Hagedorn spectrum and equation of state of Yang-Mills theories}

\ShortTitle{Hagedorn spectrum and equation of state of Yang-Mills theories}

\author{Michele~Caselle, \speaker{Alessandro~Nada}, Marco~Panero\\
        Department of Physics, University of Turin and INFN, Turin\\
        Via Pietro Giuria 1, I-10125 Turin, Italy\\
        E-mail: \email{caselle@to.infn.it}, \email{anada@to.infn.it}, \email{marco.panero@unito.it}}

\abstract{We present a novel lattice calculation of the equation of state of SU(2) Yang-Mills theory in the confining phase. 
We show that a gas of massive, non-interacting glueballs describes remarkably well the results, provided that a bosonic closed-string model is used to derive an exponentially growing Hagedorn spectrum for the heavy glueball states with no free parameters. 
This effective model can be applied to SU(3) Yang-Mills theory and the theoretical prediction agrees nicely with the lattice results reported by Bors\'anyi et al. in JHEP 07 (2012) 056.}

\FullConference{The 33rd International Symposium on Lattice Field Theory\\
                 14 -18 July  2015\\
                 Kobe International Conference Center, Kobe, Japan}

\begin{document}

\section{Introduction}

The lattice study of pure-glue gauge theories gives interesting information on the behavior of non-Abelian gauge theories, at a fraction of the typical computational costs of lattice QCD, and offers the possibility to compare the results of Monte Carlo numerical integrations with analytical calculations.
Here we present a summary of our recent work~\cite{Caselle:2015tza} in which, following this line of research, we studied the equilibrium thermodynamics of SU(2) pure-glue theory in its confining phase, i.e. for temperatures $T$ below the critical deconfinement temperature $T_c$.
We show that the equation of state in the $T<T_c$ region can be modeled as a gas of non-interacting relativistic glueballs, provided that the contribution of heavier glueball states is described in terms of a bosonic string model.
This work can be considered as a generalization of the study presented in ref.~\cite{Meyer:2009tq} for SU(3) Yang-Mills theory, later extended to SU($N$) theories in 2+1 spacetime dimensions~\cite{Caselle:2011fy}. Related ideas have also been discussed in refs.~\cite{Buisseret:2011fq, Megias:2014bfa}. The motivation to focus on the theory with $N=2$ color charges is that it provides a crucial cross-check for the string model, as it admits only states with charge conjugation $C=+1$ and it is characterized by a second-order deconfinement transition.
Furthermore we show that similarly good agreement (with no free parameters) is also found for the SU($3$) theory, using lattice data computed in ref.~\cite{Borsanyi:2012ve}. 

\section{Thermodynamics on the lattice}

A quantity of major phenomenological interest in finite-temperature field theory is the pressure $p$, which in the thermodynamic limit $V \to \infty$ equals the opposite of the free-energy density$f$:
\begin{equation}
\label{pressuretdyn}
p = -\lim_{V \to \infty} f = \lim_{V \to \infty} \frac{T}{V} \ln{Z}.
\end{equation}
The pressure $p$ is related to $\Delta$, the trace of the energy-momentum tensor (also called trace anomaly)
\begin{equation}
\frac{\Delta}{T^4} = T \frac{\partial}{\partial T} \left( \frac{p}{T^4} \right).
\end{equation}
Two other quantities such as energy density and entropy density can be readily evaluated:
\begin{equation}
\epsilon = \frac{T^2}{V} \left. {\frac{\partial \ln Z}{\partial T}}\right|_V = \Delta + 3p, \qquad s = \frac{\epsilon}{T} + \frac{\ln Z}{V} = \frac{\Delta + 4p}{T}.
\end{equation}
\\
The SU(2) Yang-Mills gauge theory is regularized on a four-dimensional hypercubic lattice $\Lambda$ of spacing $a$ and hypervolume $a^4(N_s^3 \times N_t)$, with periodic boundary conditions on all directions.
The temperature of the theory, according to thermal field theory, is the inverse of the shortest (temporal) size, i.e. $T=1/(aN_t)$. 
In this work, variations of the temperature are performed by changing the lattice spacing $a$ (which is a function of the coupling) while keeping $N_t$ fixed.
The lattice version of the action is set to be the standard Wilson action
\begin{equation}
\label{wilson_action}
\SW = -\frac{2}{g^2} \sum_{x \in \Lambda} \sum_{0 \le \mu < \nu \le 3} \Tr U_{\mu\nu} (x)
\end{equation}
where $g$ is the bare lattice coupling (which is related to the Wilson parameter $\beta=4/g^2$) and $U_{\mu\nu}(x)$ denotes the plaquette from the site $x$ in the $(\mu,\nu)$ plane.

The dynamics of the lattice system is defined by the partition function
\begin{equation}
Z = \int \prod_{x \in \Lambda} \prod_{\mu = 0}^{3} \dd U_\mu(x) e^{-\SW}
\end{equation}
so that the expectation value of a generic, gauge-invariant quantity $A$ is given by
\begin{equation}
\label{vev}
\langle A \rangle = \frac{1}{Z} \int \prod_{x \in \Lambda} \prod_{\mu = 0}^{3} \dd U_\mu(x)\, A \, e^{-\SW}.
\end{equation}
Any expectation value is estimated numerically via Monte Carlo numerical integration averaging on a large set of configurations generated by a mix of ``heat-bath'' and ``overrelaxation'' algorithms.

Thermodynamic quantities can be obtained from plaquette expectation values by the ``integral method''~\cite{Engels:1990vr}: the pressure (with respect to the value it takes at $T=0$) is given by
\begin{equation}
\label{pressure1}
p = \frac{6}{a^4} \int_{0}^{\beta} \dd \beta' \left( \langle \Up \rangle_T -  \langle \Up \rangle_0 \right)
\end{equation}
where $\langle \Up \rangle_T$ denotes the average plaquette at a generic temperature $T$. The integrand in eq.~(\ref{pressure1}) is closely related to the trace anomaly $\Delta$, since:
\begin{equation}
\label{trace1}
\Delta = \frac{6}{a^4} \frac{\partial \beta}{\partial \ln a} \left( \langle \Up \rangle_0 -  \langle \Up \rangle_T \right),
\end{equation}
where $\partial \beta / \partial (\ln a)$ can be readily evaluated from the scale setting.

\section{Scale setting}

\begin{figure}
\begin{center}
\includegraphics*[width=0.65\textwidth]{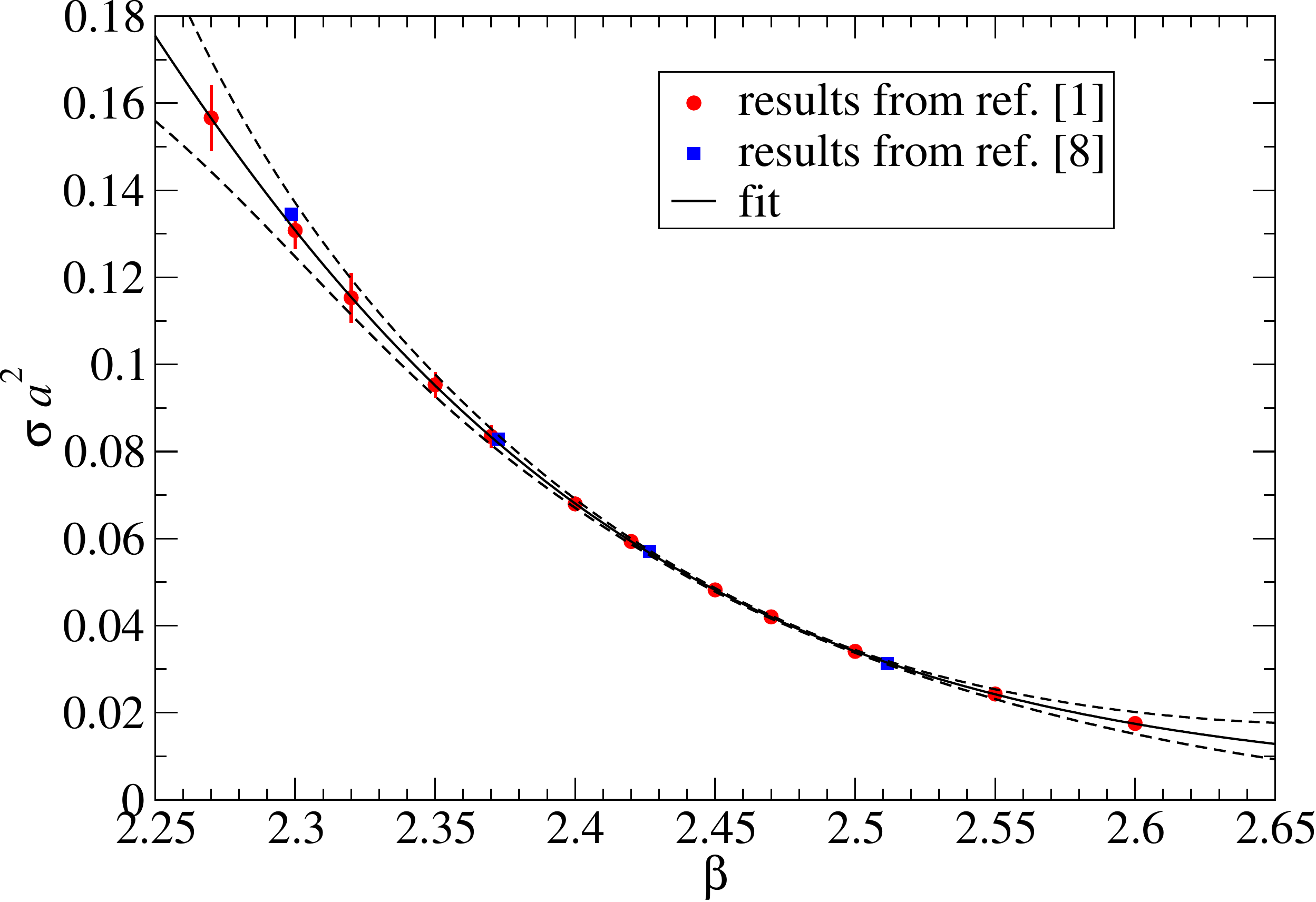}
\caption{\label{fig:scalesetting} Values of string tension in units of $a$ obtained from our lattice simulations are showed along with those reported in ref.~\cite{Lucini:2003zr}. The solid black curve shows the interpolation to the functional form in eq.~(\protect\ref{betaform}) with the associated uncertainties.}
\end{center}
\end{figure}

In order to determine the temperature reliably at a certain value of $\beta$, a precise scale setting of the theory is mandatory.
We computed non-perturbatively the zero-temperature Polyakov loop correlation function, denoted as $G(r)$, for different values of $r$ and $\beta$, using the multilevel algorithm~\cite{Luscher:2001up} on $32^4$ lattices.
The interquark potential $V(r)$ was extracted from $V(r) = - [ \ln G(r) ]/ (aN_t) $ and fitted to the functional form
\begin{equation}
\label{cornell}
a V = a \sigma r + a V_0 - \frac{\pi a}{12 r}
\end{equation}
using the tree-level improved definition of the distance $r$~\cite{Necco:2001xg} to obtain the string tension $\sigma$ in units of the squared inverse lattice spacing. As a final step, we performed a polynomial interpolation for the logarithm of the string tension for different values of the Wilson parameter $\beta$ 
\begin{equation}
\label{betaform}
\log (\sigma a^2) = \sum_{j=0}^{\npar-1} a_j (\beta-\beta_0)^j
\end{equation}
with $\npar = 4$ and $\beta_0 = 2.35$. 
The result, which models the relation between $a$ and $\beta$, is shown in fig.~\ref{fig:scalesetting}, and allows an accurate determination of the temperature for a large range of $\beta$.

\section{Numerical results and comparison with a bosonic string model}

\begin{figure}
\begin{center}
\includegraphics*[width=0.7\textwidth]{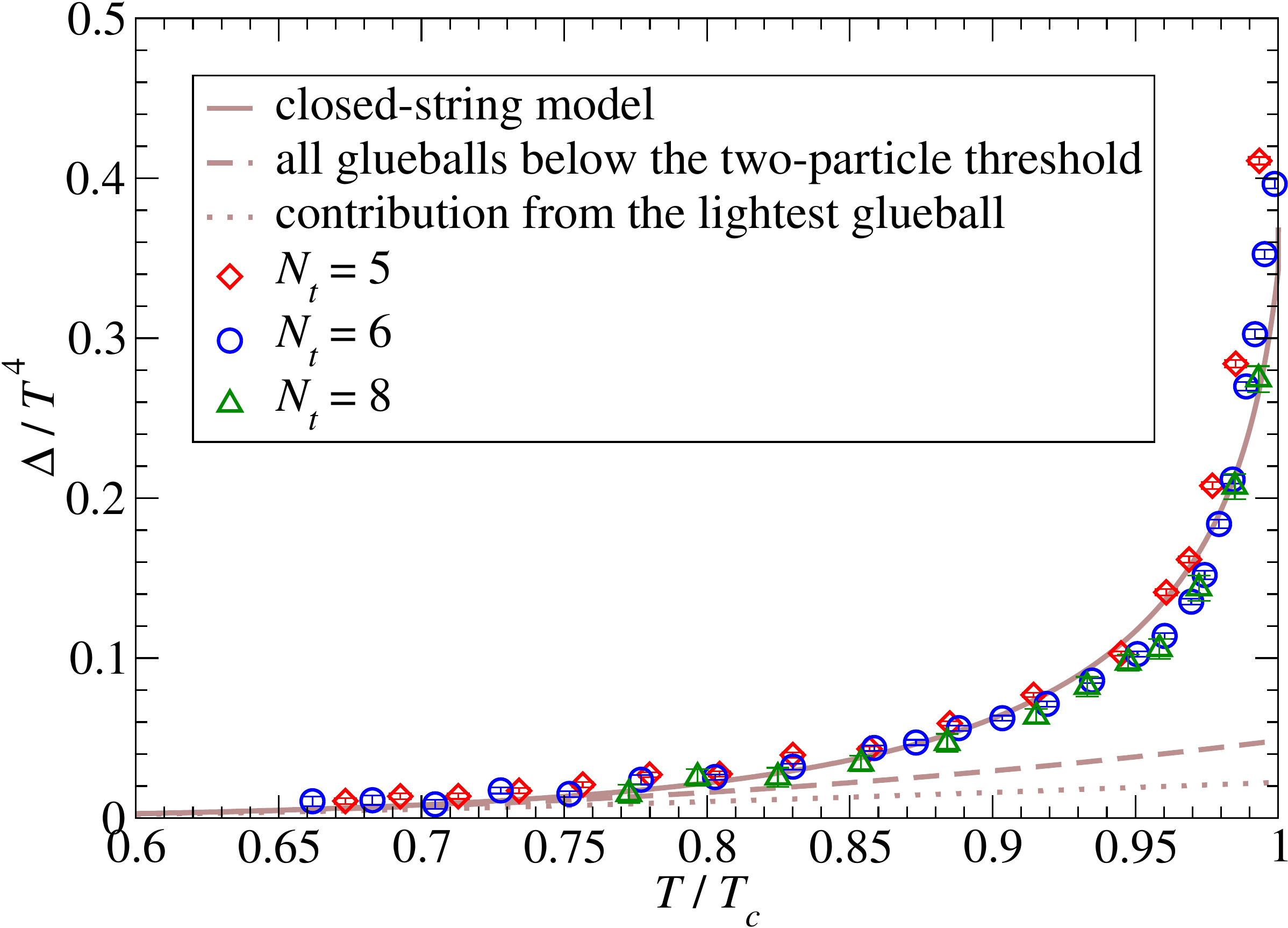}
\caption{\label{fig:su2_trace} Comparison between our lattice results for the trace anomaly in $\SU(2)$ Yang-Mills theory from simulations with different $N_t$ and the behavior expected for a gas of free, massive glueballs. The dotted line corresponds to the contribution of the lightest state only, with quantum numbers $J^{PC}=0^{++}$, while the dashed line includes all the low lying glueballs with masses lower than $2m_{0^{++}}$. The solid line includes also the contribution from high-lying states, described by a bosonic string model.}
\end{center}
\end{figure}

The main part of our numerical study of the SU(2) Yang-Mills theory is focused on the equation of state in the confining phase of the theory.
The results for $\Delta/T^4$ were obtained via Monte Carlo simulations on lattices with different temporal extents, keeping the aspect ratio $N_s/N_t$ large enough to avoid finite-volume effects.
The results are showed in figure~\ref{fig:su2_trace} and are plotted against $T/T_c$ using $T_c/\sqrt{\sigma} = 0.7091(36)$ from ref.~\cite{Lucini:2003zr}.
The only physical degrees of freedom of the theory in the $T<T_c$ region are massive glueballs: it is reasonable to assume that such states are weakly interacting with each other\footnote{The expectation that glueballs are weakly interacting is borne out of theoretical arguments in the large-$N$ limit, but lattice results indicate that these expectations are surprisingly accurate even for the theories with $N=3$ or $N=2$ color charges~\cite{Lucini:2012gg, Panero:2012qx}.} and thus the system can be modelled with good approximation as a free, relativistic Bose gas.
The trace anomaly of the latter is given by
\begin{equation}
\Delta = \frac{m^3T}{2 \pi^2} \sum_{n=1}^\infty \frac{K_1 \left( nm/T \right)}{n},
\label{B2}
\end{equation}
and using asymptotic expressions for the modified Bessel function we have
\begin{equation}
\Delta \simeq m \left( \frac{T m}{2 \pi} \right)^{3/2} \sum_{n=1}^\infty \frac{\exp \left(-nm/T \right)}{n^{3/2}} \left( 1 + \frac{3T}{8nm} \right).
\label{exp2}
\end{equation}

In figure~\ref{fig:su2_trace} the contributions due to the lightest glueball only (the $0^{++}$ state) and to all the states with mass $m<2m_{0^{++}}$ (taken from the spectrum calculated in ref.~\cite{Teper:1998kw}) are shown.
The most striking feature of the figure is the large mismatch between the glueball gas prediction and lattice data for $T$ close to $T_c$. To address this mismatch, we assume that the density of heavier glueball states is described by a Hagedorn spectrum: indeed, only an exponentially increasing spectrum can account for the exponential dependence in eq.~(\ref{exp2}).
In particular, a Hagedorn-like spectrum arises if we model glueball states as thin closed color flux tubes that can be described in terms of closed bosonic strings~\cite{Isgur:1984bm, Johnson:2000qz}.
Specifically, such closed bosonic string model leads to a spectral density (see ref.~\cite[appendix]{Caselle:2011fy} for a derivation)
\begin{equation}
 \hat{\rho} (m) = \frac{1}{m} \left( \frac{2\pi \THagedorn}{3m}\right)^{3} \exp \left( m/\THagedorn \right)
\end{equation}
where the only free parameter is the Hagedorn temperature $\THagedorn$~\cite{Hagedorn:1965st}.
If the effective action governing the string model is identified with the Nambu-Got{\={o}} action \cite{Nambu:1974zg,Goto:1971ce}, then $\THagedorn$ is fixed and its value is
\begin{equation}
\label{Hagedorntemp}
 \THagedorn = \TNG =\sqrt{\frac{3 \sigma}{2\pi}} \simeq 0.691 \sqrt{\sigma}.
\end{equation}
For SU(2) Yang-Mills theory, however, the deconfinement transition is second order and the Hagedorn temperature coincides with $T_c$, so that no determination of $\THagedorn$ is required.
Whether the transition is continuous or not, the contribution of the complete glueball spectrum for a thermodynamic quantity such as the trace anomaly $\Delta$ can be written as
\begin{equation}
\label{glueballgas}
\Delta(T) = \sum_{m_i < \mthreshold} (2J+1) \Delta (m_i,T) + n_C \int_{\mthreshold}^{\infty} \dd m^\prime \hat{\rho}(m^\prime) \Delta(m^\prime,T),
\end{equation}
where the first term includes the contribution of low-lying states (whose masses are taken from independent lattice calculations) up to a threshold chosen as $\mthreshold \equiv 2m_{0^{++}}$, and the second term approximates the contribution of heavier states via the bosonic string spectral density
The (pseudo)real nature of the representations of the SU(2) Lie group allows only glueball states with quantum number $C=+1$: thus the multiplicity factor $n_C$ is set to be $1$.
The final result can be seen in figure~\ref{fig:su2_trace}: lattice data and the bosonic string model prediction are in remarkable agreement.

\begin{figure}
\begin{center}
\includegraphics*[width=0.7\textwidth]{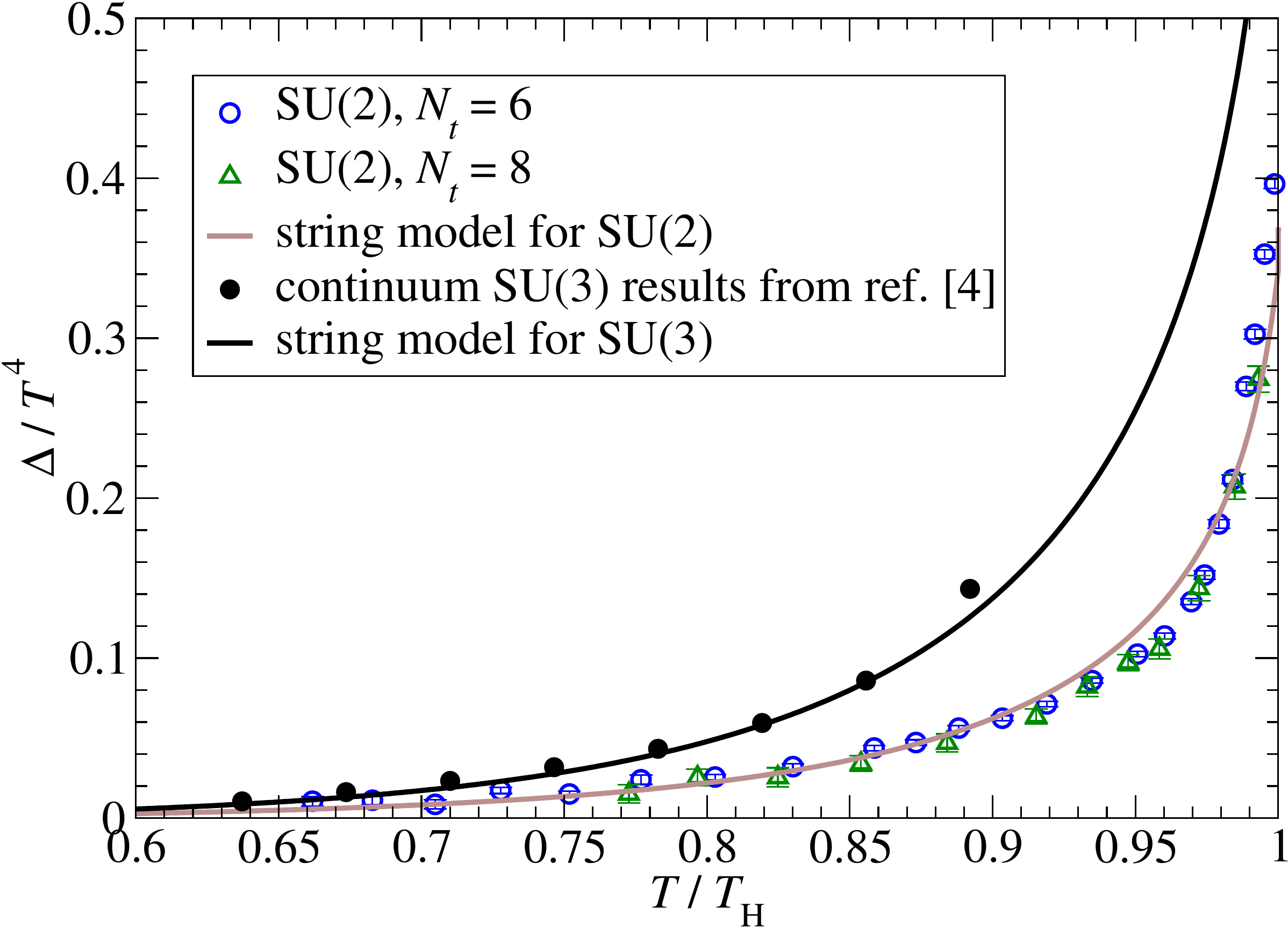}
\caption{\label{fig:su2su3} Comparison between the prediction of a massive-glueball gas, including the contribution from states modelled by a closed Nambu-Got\={o} string model, like in eq.~(\protect\ref{glueballgas}), and continuum-extrapolated data obtained in ref.~\cite{Borsanyi:2012ve} for $\SU(3)$ Yang-Mills theory, as a function of $T/\THagedorn$. Our data and predictions for the $\SU(2)$ theory are also showed for comparison.}
\end{center}
\end{figure}

We tested this model also for the SU(3) theory, using the results for the equation of state from ref.~\cite{Borsanyi:2012ve} and the glueball spectrum from ref.~\cite{Meyer:2004gx}. The differences from the $N=2$ case are:
\begin{itemize}
\item SU(3) Yang-Mills theory allows for both $C=\pm 1$ states, thus the factor $n_C$ in eq.~(\ref{glueballgas}) is set to 2 in order to account for this charge-conjugation multiplicity;
\item the deconfinement transition is of first order and the value of the Hagedorn temperature is fixed by the Nambu-Got{\={o}} prediction, see eq.~(\ref{Hagedorntemp}).
\end{itemize}
The resulting curve for the trace is showed in figure~\ref{fig:su2su3} as a function of $T/\THagedorn$, along with the results for SU(2).
The parameter $T_c/\sqrt{\sigma} = 0.629(3)$ from ref.~\cite{Boyd:1996bx} was used to plot the data; this value is consistent with the recent result $r_0T_c = 0.7457(45)$ taken from ref.~\cite{Francis:2015lha} and combined with $r_0 \sqrt{\sigma} = 1.192(10)$ from ref.~\cite{Guagnelli:1998ud}.
Excellent agreement with the effective string prediction is also found for SU(3) Yang-Mills theory, and the contribution of $C=-1$ states in the spectrum is crucial.
We remark that the non-interacting glueball gas predictions for both $N=2$ and $N=3$ colors do not depend on any free parameters: furthermore, these findings are in agreement with previous results in $D=2+1$ spacetime dimensions for SU($N=2,...,6$) theories~\cite{Caselle:2011fy} and can be considered a generalization of those.


\begin{thebibliography}{99}
  
  \bibitem{Caselle:2015tza}
  M.~Caselle, A.~Nada and M.~Panero,
  {\it {Hagedorn spectrum and thermodynamics of SU(2) and SU(3) Yang-Mills theories}},
  {\em JHEP} {\bf 07} (2015) 143,
  [\href{http://xxx.lanl.gov/abs/1505.01106}{{\tt arXiv:1505.01106}}].
  

  \bibitem{Meyer:2009tq}
  H.~B. Meyer,
  {\it {High-Precision Thermodynamics and Hagedorn Density of
  States}},
  {\em Phys.Rev.} {\bf D80} (2009) 051502,
  [\href{http://xxx.lanl.gov/abs/0905.4229}{{\tt arXiv:0905.4229}}].
  
  \bibitem{Caselle:2011fy}
  M.~Caselle et al.,
  {\it {Thermodynamics of SU(N) Yang-Mills theories in 2+1 dimensions I - The
  confining phase}},
  {\em JHEP} {\bf 1106} (2011) 142,
  [\href{http://xxx.lanl.gov/abs/1105.0359}{{\tt arXiv:1105.0359}}].
  
  \bibitem{Buisseret:2011fq}
  F.~Buisseret and G.~Lacroix,
  {\it {Comments on Yang-Mills thermodynamics, the Hagedorn spectrum and the gluon gas}},
  {\em Phys.Lett.} {\bf B705} (2011) 405-409,
  [\href{http://xxx.lanl.gov/abs/1105.1092}{{\tt arXiv:1105.1092}}].

  \bibitem{Megias:2014bfa}
  E.~Meg{\'i}as, E.~Ruiz Arriola and L.~L.~Salcedo,
  {\it {Polyakov loop spectroscopy in the confined phase of gluodynamics and QCD}},
  {\em Nucl.Part.Phys.Proc.} {\bf 258-259} (2015) 201,
  [\href{http://xxx.lanl.gov/abs/1409.0773}{{\tt arXiv:1409.0773}}].
  
  \bibitem{Borsanyi:2012ve}
  S.~Bors{\'a}nyi et al.,
  {\it {Precision SU(3) lattice thermodynamics for a large temperature range}},
  {\em JHEP} {\bf 1207} (2012) 056,
  [\href{http://xxx.lanl.gov/abs/1204.6184}{{\tt arXiv:1204.6184}}].
  
  \bibitem{Engels:1990vr}
  J.~Engels et al.,
  {\it  {Nonperturbative thermodynamics of SU(N) gauge theories}},
  {\em Phys.Lett.} {\bf B252} (1990) 625--630.
  
  \bibitem{Luscher:2001up}
  M.~L{\"u}scher and P.~Weisz,
  {\it {Locality and exponential error reduction in numerical lattice gauge theory}},  
  {\em JHEP} {\bf 0109} (2001) 010,
  [\href{http://xxx.lanl.gov/abs/hep-lat/0108014}{{\tt hep-lat/0108014}}].
  
  \bibitem{Necco:2001xg}
  S.~Necco and R.~Sommer,
  {\it {The N(f) = 0 heavy quark potential from short to intermediate distances}},
  {\em Nucl.Phys.} {\bf B622} (2002) 328--346,
  [\href{http://xxx.lanl.gov/abs/hep-lat/0108008}{{\tt hep-lat/0108008}}].
  
  \bibitem{Lucini:2003zr}
  B.~Lucini, M.~Teper and U.~Wenger,
  {\it {The High temperature phase transition in SU(N) gauge theories}},
  {\em JHEP} {\bf 0401} (2004) 061,
  [\href{http://xxx.lanl.gov/abs/hep-lat/0307017}{{\tt hep-lat/0307017}}].
  
  \bibitem{Lucini:2012gg}
  B.~Lucini and M.~Panero,
  {\it {SU(N) gauge theories at large N}},
  {\em Phys.Rept.} {\bf 526} (2013) 93,
  [\href{http://arxiv.org/abs/1210.4997}{{\tt arXiv:1210.4997}}].
  
  \bibitem{Panero:2012qx}
  M.~Panero,
  {\it {Recent results in large-N lattice gauge theories}},
  {\em PoS} {\bf Lattice 2012} (2012) 010,
  [\href{http://arxiv.org/abs/1210.5510}{{\tt arXiv:1210.5510}}].

  \bibitem{Teper:1998kw}
  M.~J. Teper,
  {\it {Glueball masses and other physical properties of SU(N) gauge theories in D = (3+1): A Review of lattice results for theorists}},
  \href{http://xxx.lanl.gov/abs/hep-th/9812187}{{\tt hep-th/9812187}}.
  
  \bibitem{Isgur:1984bm}
  N.~Isgur and J.~E. Paton,
  {\it {A Flux Tube Model for Hadrons in QCD}},
  {\em Phys.Rev.} {\bf D31} (1985) 2910.
  
  \bibitem{Johnson:2000qz}
  R.~W.~Johnson and M.~J.~Teper, 
  {\it {String models of glueballs and the spectrum of SU(N) gauge theories in (2+1)-dimensions}},
  {\em Phys.Rev.} {\bf D66} (2002) 036006,
  [\href{http://xxx.lanl.gov/abs/hep-ph/0012287}{{\tt hep-ph/0012287}}].
  
  \bibitem{Hagedorn:1965st}
  R.~Hagedorn, 
  {\it {Statistical thermodynamics of strong interactions at
  high-energies}},
  {\em Nuovo Cim.Suppl.} {\bf 3} (1965) 147--186.
  
  \bibitem{Nambu:1974zg}
  Y.~Nambu, 
  {\it {Strings, Monopoles and Gauge Fields}},
  {\em Phys.Rev.} {\bf D10} (1974) 4262.

  \bibitem{Goto:1971ce}
  T.~Got{\={o}}, 
  {\it {Relativistic quantum mechanics of one-dimensional mechanical continuum and subsidiary condition of dual resonance model}},
  {\em Prog.Theor.Phys.} {\bf 46} (1971) 1560--1569.
  
  \bibitem{Meyer:2004gx}
  H.~B. Meyer, {\it {Glueball regge trajectories}},
  \href{http://xxx.lanl.gov/abs/hep-lat/0508002}{{\tt hep-lat/0508002}}.

  \bibitem{Boyd:1996bx}
  G.~Boyd et al.,
  {\it {Thermodynamics of SU(3) lattice gauge theory}},  {\em Nucl.Phys.} {\bf B469}
  (1996) 419--444,
  [\href{http://xxx.lanl.gov/abs/hep-lat/9602007}{{\tt hep-lat/9602007}}].
  
  \bibitem{Francis:2015lha}
  A.~Francis et al.,
  {\it {Critical point and scale setting in SU(3) plasma: An update}},
  {\em Phys.Rev.} {\bf D91} (2015) 9,  096002,
  [\href{http://xxx.lanl.gov/abs/1503.05652}{{\tt arXiv:1503.05652}}].
  
  \bibitem{Guagnelli:1998ud}
  M.~Guagnelli, R.~Sommer and H.~Wittig [ALPHA Collaboration],
  {\it {Precision computation of a low-energy reference scale in quenched lattice QCD}},
  {\em Nucl.Phys.} {\bf B535} (1998) 389, [\href{http://xxx.lanl.gov/abs/hep-lat/9806005}{{\tt hep-lat/9806005}}].

\end{thebibliography}
\end{document}